\documentclass[11pt,a4paper]{article}
\usepackage{amsfonts}
\usepackage{graphicx}
\usepackage{amsmath}
\usepackage{hyperref}
\usepackage{enumerate}

\RequirePackage{mathrsfs} \RequirePackage[sc]{mathpazo}
\RequirePackage{wasysym} \RequirePackage{setspace}
\makeatletter \@addtoreset{equation}{section}

\newcommand{\be}{\begin{equation}}
\newcommand{\ee}{\end{equation}}
\newcommand{\bea}{\begin{eqnarray}}
\newcommand{\eea}{\end{eqnarray}}
\begin{document}

\title{\textbf{\ Embedding Fractional Quantum Hall Solitons in}\\
\textbf{M-theory Compactifications} }
\author{A. Belhaj$^{1,2,5}$\thanks{%
belhaj@unizar.es}, N-E. Fahssi$^{1,3,5}$\thanks{%
fahssi@uh2m.ac.ma}, E.H Saidi$^{1,5}$\thanks{%
h-saidi@fsr.ac.ma}, A. Segui$^{4}$\thanks{%
segui@unizar.es}\hspace*{-15pt} \\
\\
{\small $^{1}$Lab Phys Hautes Energies, Modelisation et Simulation, Facult%
\'{e} des Sciences, Rabat, Morocco} \\
{\small $^{2}$Centre National de l'\'Energie, des Sciences et des Techniques
Nucl\'eaires, Rabat, Morocco } \\
{\small $^{3}$D{\'e}partement de Math{\'e}matiques, Facult{\'e} des Sciences
et Techniques, Mohammedia, Morocco} \\
{\small $^{4}$Departamento de F\'isica Te\'orica, Universidad de Zaragoza,
E-50009-Zaragoza, Spain}\\
{\small $^{5}$Groupement National de Physique des Hautes Energies, Si\`{e}ge
focal: FSR, Rabat, Morocco }}
\maketitle

\begin{abstract}
We engineer U(1)$^n $ Chern-Simons type theories describing fractional
quantum Hall solitons (QHS) in 1+2 dimensions from \mbox{M-theory}
compactified on eight dimensional hyper-K\"{a}hler manifolds as target space
of $N=4$ sigma model. Based on M-theory/Type IIA duality, the systems can be
modeled by considering D6-branes wrapping intersecting Hirzebruch surfaces $%
F_0$'s arranged as  $ ADE$  Dynkin Diagrams and interacting with higher dimensional
R-R gauge fields. In the case of finite Dynkin quivers, we  recover well
known values of the filling factor observed experimentally including  Laughlin, Haldane and  Jain  series.  \newline
\end{abstract}

\newpage

\section{Introduction}

Recently, efforts have been devoted to study connections between the quantum
theory of condensed matter physics and higher dimensional supergravity
models embedded in 10D type II superstrings and 11D M-theory \cite%
{-1,0,BR,Rey}. In particular, the fractional Quantum Hall Effect (QHE) has
been subject to some interest not only because of its experimental results,
including graphene \cite{GRA}, but also from its connection with the recent
developments in brane physics using Anti de Sitter/conformal field theory
(AdS/CFT) correspondence \cite{Ma} and string theory compactifications \cite%
{BS}.

The first proposed series of the fractional quantum states was given
by
Laughlin and they are characterized by the filling factor $\nu _{L}=\frac{1}{%
m}$ where $m$ is an even integer for a boson electron and an odd integer for
a fermionic electron \cite{L,H}. At low energy, this model can be described
by a 3-dimensional U(1) Chern-Simons theory coupled to an external
electromagnetic field $\tilde{A}$ with the following effective action
\begin{equation}
S_{CS}=-\frac{m}{4\pi }\int_{\mathbb{R}^{1,2}}A\wedge dA+\frac{q}{2\pi }%
\int_{\mathbb{R}^{1,2}}{\tilde{A}}\wedge dA  \label{sc}
\end{equation}%
where $A$ is the dynamical gauge field and $q$ is the charge of the electron
\cite{wen}. It turns out that this system can be modeled using solitonic
D-branes of type II superstrings with a NS-NS \mbox{B-field} \cite%
{BBST,FLRT,BS,BJLL}. When the B-field is turned on, a non
commutative geometry description can be also used
\cite{ES}.

Following the Susskind approach and looking for extended
constructions, it is not difficult to see that the most general
fractional quantum Hall systems including (\ref{sc}) is given by the
following abelian effective theory
\begin{equation}
\begin{tabular}{ll}
$S\sim \frac{1}{4\pi }\int \sum_{i,j}K_{ij}A^{i}\wedge dA^{j}+2\sum_{i}q_{i}%
\tilde{A}\wedge dA^{i}$ &
\end{tabular}
\label{hd}
\end{equation}%
where now $K_{ij}$\ is a real, symmetric and invertible matrix ($\det K\neq 0
$); and $q_{i}$\ is a vector of charges. The apparition of the $K_{ij%
}$\ matrix and the $q_{i}$\ charge vector in this effective field
action are very suggestive in the sense that, besides their Lie
algebra interpretation, they also capture\textrm{\ }the property to
embed gauge theory in type II superstrings and, via string
dualities, in M-theory
compactifications. Moreover, by integrating over the gauge fields $%
A^{i}=dx^{\mu }A_{\mu }^{i}$\ in the same way as in Susskind model, we get
the following filling factor
\begin{equation}
\begin{tabular}{ll}
$\nu =q_{i}K_{ij}^{-1}q_{j}$ &
\end{tabular}
\label{factor}
\end{equation}%
letting understand that eq(\ref{factor}) may be also thought of as
giving a \emph{unified description} of several kinds of FQH series
including\ Laughlin, Haldane, Jain and hierarchical ones
\textrm{\cite{ES,SE}. }An appropriate choice of $K_{ij}$\ and
$q_{i}$\ leads to a particular filling factor.

In this letter we discuss Fractional Quantum Hall Solitons (QHS)
described by gauge quivers in 1+2 associated with the class of
Kac-Moody Lie algebras having $\det K\neq 0$. This class includes
the subset of ordinary Lie algebras classified by Cartan and a
sector in the so called indefinite
subset \cite{V}. Notice that dealing with a general form of $K_{ij}$\textrm{%
\ }would be interesting; but this requires considering Borcherds
algebras \cite{BOR}. However the geometric interpretation of these
exotic Borcherds symmetries go beyond the usual intersecting
\mbox{2-(4-)} cycles as it happens generally in the embedding of
quiver gauge theories in superstring compactifications.

As such in our realization, the matrix $K_{ij}$\ will be identified with the
Cartan matrix and its extensions to those having $\det K\neq 0$ with
indefinite sign such as hyperbolic algebras considered in \cite{MAL}.\textrm{%
\ }The corresponding models are obtained from a direct compactification of %
\mbox{M-theory} on a real eight dimensional manifold (complex
4-dimensional hyper-K\"{a}hler manifold). The geometry is realized
explicitly as a
cotangent bundle over a collection of intersecting Hirzebruch surfaces%
\footnote{%
The Hirzebruch surface ($F_{0}$) is defined by a trivial fibration of $%
\mathbb{CP}^{1}$ over $\mathbb{CP}^{1}$; it is a particular del Pezzo
surface ($dP_{1}$) with toric realization given by a rectangle.} ($F_{0}$)
arranged as Dynkin diagrams. Notice by the way that $F_{0}$ is a complex
compact surface with simpler homology properties; general realizations are
also possible; for instance by considering del Pezzo surfaces \cite{AS} \`{a}
la F-theory-GUT \cite{W,BV} where the $K_{ij}$\ is encoded in the degeneracy
of the elliptic fiber\footnote{%
It would be interesting to deepen this issue as it includes the remarkable
case $\det K_{ij}=0$ which is associated with affine singularities \cite%
{BV,BG} and also space-time conformal symmetry \cite{MAL}.} of the
elliptically fibered CY 4-folds (see also the comment made in the conlusion
section).\textrm{\ }More general extensions could be also done for toric
varieties in which the matrix $K_{ij}$ is identified with Mori matrices.%

Moreover, based on M-theory/Type IIA duality, we give $N=1$ supersymmetric $%
\mbox{U(1)}^{n}$ Chern-Simons type theories describing 3-dimensional QHS
using D6-branes wrapping intersecting $F_{0}$'s in the presence of higher
dimensional R-R gauge fields. In the case of finite Dynkin quivers, we
geometrically recover some very known values observed experimentally
including the Jain's series. We expect to have similar results for
hyperbolic quiver models with negative filling fractions describing holes.%

The organization of this work is as follows. First, we give the
M-theory background as target space of $N=4$ sigma model. In section
3, we derive QHS in 1+2 dimensions using M-theory/Type IIA duality.
In section 4, we compute the filling factor for $A_{n}$ quivers. For
a particular choice of the vector charge, we can recover the Jain's
series. Then we extend the analysis to $DE$ quivers in section 5.
Our conclusion and comments are given in section 6.

\section{ M-theory background}

In this section we give the geometric background that we will use to derive
QHS in 1+2 dimensions from M-theory compactified on a special class of eight
dimensional manifold $M_{8}$ namely a 4-dimensional complex hyper-K\"{a}hler
manifold
\begin{equation}
\mathbb{R}^{1,2}\times M_{8}
\end{equation}%
leading to a $N=2$ supersymmetric low energy effective Chern-Simons gauge
theory in 3D space time. For a type II stringy interpretation, $M_{8}$ will
be viewed as a circle fibration over a real 7-dimensional base manifold $%
M_{7}$. Using M-theory/Type IIA duality, the base $M_{7}$ can be
identified with a local type IIA geometry. A nice way to describe
$M_{8}$ is to use the so-called hyper-K\"{a}hler quotient studied in
\cite{GVW} engineered by considering a two-dimensional U(1)$^{r}$
sigma model with eight supercharges and $r+2$ hypermultiplets. There
is a SU$(r+2)$ global symmetry
under which the hypermultiplets transform in the fundamental representation $%
r+2$. Thus, $M_{8}$ is defined by the following D-flatness condition
\begin{equation}
\sum_{i=1}^{r+2}Q_{i}^{a}[\phi _{i}^{\alpha }{\bar{\phi}}_{i\beta }+\phi
_{i}^{\beta }{\bar{\phi}}_{i\alpha }]=\vec{\xi}_{a}\vec{\sigma}_{\beta
}^{\alpha },\;\;a=1,\ldots ,r  \label{sigma4}
\end{equation}%
where $Q_{i}^{a}$ is a matrix charge which will be identified here with the
extended Cartan matrices \cite{V}. $\phi _{i}^{\alpha }$'s ($\alpha =1,2)$
denote the component field doublets of each hypermultiplets ($i=1\ldots
,r+2) $. $\vec{\xi}_{a}$ are the Fayet-Iliopolos (FI) 3-vector couplings
rotated by SU(2) symmetry, and $\vec{\sigma}_{\beta }^{\alpha }$ are the
traceless $2\times 2$ Pauli matrices. Performing SU(2) R-symmetry
transformations $\phi ^{\alpha }=\varepsilon ^{\alpha \beta }\phi _{\beta
},\;\overline{\phi ^{\alpha }}=\overline{\phi }_{\alpha },\;\varepsilon
_{12}=\varepsilon ^{21}=1$ and replacing the Pauli matrices by their
expressions, the identities (\ref{sigma4}) can be split as follows%
\begin{equation}
\begin{tabular}{llll}
$\sum\limits_{i=1}^{r+2}Q_{i}^{a}(|\phi _{i}^{1}|^{2}-|\phi _{i}^{2}|^{2})$
& $=$ & $\xi _{a}^{3}$ &  \\
$\sum\limits_{i=1}^{r+2}Q_{i}^{a}\phi _{i}^{1}\overline{\phi }_{i}^{2}$ & $=$
& $\xi _{a}^{1}+i{\xi ^{2}}_{a}$ &  \\
$\sum\limits_{i=1}^{r+2}Q_{i}^{a}\phi _{i}^{2}\overline{\phi }_{i}^{1}$ & $=$
& $\xi _{a}^{1}-i{\xi ^{2}}_{a}$ & .%
\end{tabular}%
\end{equation}%
Up to some technical details, the general solution of these equations can be
viewed as the cotangent bundle\footnote{$F_{0}$ is the blow up of $CP^{2}$
at a point; it is just $dP_{1}$ the leading term of the del Pezzo surfaces $%
dP_{n}$ given by the blow ups of $CP^{2}$ at $n$ points with $n\leq 8$. It
would be interesting to extend the present construction to intersecting $%
dP_{n}$s.} over a collection of $r$ intersecting $F_{0}\sim \mathbb{CP}%
^{1}\times \mathbb{CP}^{1}$ arranged as Dynkin diagrams of Kac-Moody Lie
algebras \cite{1,2}. This result, which generalizes the $N=2$ scenario
dealing with the ALE spaces in which appear only intersecting $\mathbb{CP}%
^{1}$'s, leads to a nice correspondence between the root lattice of
Kac-Moody Lie algebras and the set of $F_{0}$'s forming a basis of the
cohomology space of four-cycles $H_{4}(M_{8},\mathbb{Z})$. This connection
can be supported by the intersection theory of complex surfaces inside $%
M_{8} $. In fact, the self-intersection of the zero section in the cotangent
bundle of $F_{0}$ is equal to its minus Euler number, i.e. $-4$. Assuming
that $F_{0}^{i}$ intersects $F_{0}^{i+1}$ at two points, which can be
supported by the fact that each $\mathbb{CP}^{1}$ inside $F_{0}^{i}$
intersects just one $\mathbb{CP}^{1}$ in the next $F_{0}^{i+1}$, we get the
intersection numbers of the $F_{0}$'s
\begin{equation}
\lbrack F_{0}^{i}]\cdot \lbrack F_{0}^{i}]=-4,\qquad \lbrack F_{0}^{i}]\cdot
\lbrack F_{0}^{i+1}]=2,
\end{equation}%
with others vanishing. This means that $F_{0}^{i}$ does not intersect $%
F_{0}^{j}$ if $|j-i|>1$. Up to a multiplication factor by two, the
intersection numbers reproduce the elements of the Cartan matrices $C_{ij}$.
The compact intersecting geometry agrees with the extended Dynkin diagrams.
More specifically, with each simple root $\alpha _{i}$, we associate a
single $F_{0}^{i}$ and we have the following intersection form
\begin{equation}
\lbrack F_{0}^{i}]\cdot \lbrack F_{0}^{j}]=-2C^{ij}.  \label{Kij}
\end{equation}%
It should be interesting to write down the algebraic equations dealing with
the corresponding singularities extending the case of ALE spaces \cite{KMV}.
We expect that this could be obtained in terms of $N=4$ sigma model gauge
invariants.

\section{ QHS from M-theory/Type IIA duality}

Having specified the geometric background, we will give a M-theory QHS in
terms of effective Chern-Simons type theory with a series of U(1) gauge
fields. This internal space presents $ADE$ quiver description of QHS in 1+2
dimensions which are obtained from a direct M-theory compactification
instead of type IIA moving on $ALE\times S^{3}$ and performing a
Kaluza-Klein compactification on $S^{3}$ \cite{FLRT}. Our way provides a
different brane system dual to M-theory geometric background leading to
fractional QHS. Besides that, it will allow us to derive directly some
filling fractions which coincide with a subsequence of the celebrated Jain's
series. This can be recovered as exact values without modifying the
intersection matrix associated with the re-normalization of the inner
product between simple roots as made in \cite{BEFKSS} for the $ALE$ space in
the presence of D4-branes. Moreover, the M-theory internal space can develop
singularities of many different types and we expect that instead of $ADE$
singularties, one could in a similar way analyze other intersecting
geometries. This may lead to new quiver models based on resolving such
singularities.

Roughly, our analysis will be based on a dual type IIA local geometry in
presence of D6-branes interacting with R-R higher dimensional gauge fields.
By the use of M-theory/Type IIA duality, it is worth noting that M-theory on
$M_{8}$ is expected to be dual to Type IIA superstring on $M_{8}/\mbox{U(1)}$%
, where $\mbox U(1)$ can be identified with the M-theory circle
compactification (going from eleven dimensions to ten). A priori, there are
many ways one may follow to choose the U(1) symmetry. Here we will use the
circle actions involved in toric geometry which is a powerful tool for
studying complex manifolds in terms of simple combinatorial data of
polytopes \cite{F}. The simple example of toric varieties is the complex
plane $\mathbb{C}$. The latter admits an U(1) toric action
\begin{equation}
z\rightarrow ze^{i\theta },
\end{equation}%
which has a fixed point at $z=0$. The toric geometry of $\mathbb{C}$ can be
viewed as a circle fibred on a half line parameterized by $|z|$. The circle
determined by the action of $\theta $ shrinks at $z=0$. This realization can
be generalized easily to $\mathbb{C}^{n}$ space where we have a $T^{n}$
fibration, parameterized by the angular coordinates $\theta _{i}$, over a $n$%
-dimensional real base parameterized by $|z_{i}^{2}|$. The more interesting
compact example in toric geometry is the $\mathbb{CP}^{1}$ space admitting
an U(1) toric action having two fixed points describing respectively the
north and the south poles of the two sphere $S^{2}\sim \mathbb{CP}^{1}$. In
this way, $\mathbb{CP}^{1}$ can be viewed as an interval fibred by $S^{1}$
with zero size at the two boundaries. Using these ideas, our geometry $M_{8}$
can be viewed also as a toric space admitting four toric geometry circle
actions $\mbox{U(1)}_{base}^{2}\times \mbox{U(1)}_{fiber}^{2}$; two of them
correspond to the $F_{0}$'s base space denoted by $\mbox{U(1)}_{base}^{2}$
while the remaining ones $\mbox{U(1)}_{fiber}^{2}$ act on the fiber
cotangent directions. Dividing by one finite fiber circle action
\begin{equation}
M_{7}={\frac{M_{8}}{\mbox{U(1)}_{fiber}}},
\end{equation}%
we can obtain a 7-dimensional type IIA geometry. For instance, in
the case of two dimensional $\mbox{U(1)}^{r}$ sigma model with the
finite $A_{r}$ Cartan matrix gauge charges and $r+2$
hypermultiplets, this quotient space becomes a real cone on a
${S^{2}}$ bundle over a collection of $r$ intersecting $F_{0}$'s
arranged as Dynkin diagram of $A_{r}$ finite Lie algebras,
preserving $N=2$ supersymmetry in $2+1$ dimensions.

In the following, we will show that the dual type IIA geometry can generate $%
\mbox{U(1)}^n$ Chern-Simons type theories from D6-branes wrapped on
intersecting $F_{0}^{i}$'s and filling the 3-dimensional Minkowski space on
which QHS will reside. To obtain an effective theory of hierarchical
description QHS, let us first start with a Chern-Simons theory with a single
U(1) gauge symmetry. The corresponding geometry can obtained from a sigma
model with one U(1) gauge symmetry and $A_1$ vector charge. In this case,
the local type IIA geometry reduces to a real cone over
\begin{equation}
S^2\times F_0
\end{equation}
Using arguments similar to \cite{FLRT,M}, we can wrap a D6-brane over the
zeroth Hirzebruch surface $F_0$ to get the Chern-Simons action (\ref{sc}).
Indeed, on the seven-dimensional world-volume of each D6-brane we have U(1)
gauge symmetry. The corresponding effective theory has two parts:
\begin{equation}
S_{D6}=S_{DBI}+S_{WZ}.
\end{equation}
The DBI part is given by
\begin{equation}
S_{DBI}\sim T_6 \int d^7\sigma e^{-\phi}\sqrt{ -\det (G+2\pi F)}
\end{equation}
while the WZ action reads as
\begin{equation}
S_{WZ}\sim T_6\int_{\mathbb{R}^{1,2}\times F_0} F\wedge F\wedge C_3
\end{equation}
where $T_6$ is the brane tension and where $C_3$ is the R-R 3-form coupled
to the D2-brane of type IIA superstring. Ignoring the first terms and
integrating by part, the WZ action on the D6-brane world-volume becomes
\begin{equation}
\int_{\mathbb{R}^{1,2}\times F_0} F\wedge F\wedge C_3=-\int_{\mathbb{R}%
^{1,2}\times F_0} A\wedge F\wedge (dC)_4
\end{equation}
Now, integrating over $F_0$, we get the first Chern-Simons terms
\begin{equation}
-\frac{m}{4\pi} \int_{\mathbb{R}^{1,2}} A\wedge F
\end{equation}
where $m=\frac{1}{2\pi}\int_{F_0}(dC)_4$ which can be computed from
intersection theory of $M_8$. To couple the system to an external gauge
field, we need to introduce the RR 5-form $C_5$ which is sourced by a
D4-brane. This gauge field decomposes as follows
\begin{equation}
C_5\to \tilde A \wedge \omega
\end{equation}
where $\omega$ is a harmonic 4-form on $F_0$. In this way, the WZ term $\int
C_5 \wedge F$ on a D6-brane gives
\begin{equation}
q \int_{\mathbb{R}^{1,2}} \tilde A \wedge F
\end{equation}
where $\tilde A$ is the U(1) gauge field which can be obtained from the
dimensional reduction of the RR 5-form on $F_0$. This U(1) gauge field can
be interpreted as a magnetic external gauge field that couples to our QHS.
We can follow the same steps to construct an effective Chern-Simons gauge
theory with a series of U(1) gauge fields which is called hierarchical
description. The corresponding effective action can be obtained from a stack
of D6-branes wrapping individually intersecting $F_0$'s. Using the
intersection contributions, the action can take the same form as in (\ref{hd}%
) where now $K_{ij}$ can be identified with the intersection matrix of $F_0$%
's (\ref{ordAk}), and where $q_i$ is the vector field characterizing
a fractional quantum Hall theory associated with $ADE$ geometries. Since $K_{ij}$ and $q_{i}$, which are interpreted as order
parameters
classifying the various FQH states, are related to the filling factor \cite%
{WZ}, equation (\ref{factor}) can be solved algebraically in terms of
representation theory of the Kac-Moody Lie algebras \cite{BEFKSS}.


\section{ ADE quiver models}
 In this section, we present concrete examples of the more formal results
that we have developed above for which Chern-Simons gauge theory provides an $ADE$ effective field description. For simplicity, we will mainly consider the  $ADE$  finite quiver gauge models by introducing 4-cycles in the base which
are intersecting according to  $ADE$ Dynkin graph. We refer to these models  as $ADE$ quiver models. They constitute a very natural class of  models  in this context. It is pointed out,
though, that we in principle could consider more complicated geometries. We will restrict
ourselves to the simply laced $ADE$ ones as they allow us to extract the corresponding physics in a straightforward manner. In this case, the quadratic form (\ref{factor}) can be
written as
\begin{equation}
\nu=\frac{1}{2}\sum_{i}C_{ii}^{-1}q_i^2+\sum_{i<j}C _{ij}^{-1}q_iq_j.
\label{factora}
\end{equation}
\subsection{ $A_n$ quiver models}
As an illustration, we now consider the $A_n$ quivers  corresponding  to 
$\mbox{U(1)}^{n}$ quiver gauge theory based on finite $A_{n}$ Dynkin
diagram. In the context of Type IIA string theory, this model
appears as the world-volume of $n$ D6-branes wrapping separately
$F_{0}$'s arranged as follows
\begin{equation}
\mbox{
         \begin{picture}(20,30)(70,0)
        \unitlength=2cm
        \thicklines
    \put(0,0.2){\circle{.2}}
     \put(.1,0.2){\line(1,0){.5}}
     \put(.7,0.2){\circle{.2}}
     \put(.8,0.2){\line(1,0){.5}}
     \put(1.4,0.2){\circle{.2}}
     \put(1.6,0.2){$.\ .\ .\ .\ .\ .$}
     \put(2.5,0.2){\circle{.2}}
     \put(2.6,0.2){\line(1,0){.5}}
     \put(3.2,0.2){\circle{.2}}
     \put(-1.2,.15){$A_{n}:$}
  \end{picture}
}  \label{ordAk}
\end{equation}%
For a generic vector charge, the filling factor (\ref{factora}) is given now
by the following quadratic form
\begin{equation}
\nu (A_{n})=\frac{1}{n+1}\left( \frac{1}{2}\sum_{i=1}^{n}i(n-i+1)q_{i}^{2}+%
\sum_{i<j}i(n-j+1)q_{i}q_{j}\right) .
\end{equation}%
In the quantum Hall literature, the simplest model is related to single
layer FQH states. In this case, the components of the vector charge are
zeros except one entry which is equal 1. When taking into account that $%
q_{i}=\delta _{i,p}$ for some $p=1,\ldots ,n$, we obtain the relation
\begin{equation}
\nu (A_{n})=\nu _{p,n}=\frac{p(n+1-p)}{2(n+1)},
\end{equation}%
admitting the obvious symmetry $\nu _{p,n}=\nu _{n+1-p,n}.$ Actually,  this can be viewed as a  generlization of the result given in \cite{BS}. To see that, specializing the
computation to $p=1$ (or $n$), we get $\nu _{1,n}=\frac{n}{2(n+1)}$. Taking $%
n=2m$ (even number of $F_{0}$'s), the filling factor can be written as
\begin{equation}
\nu _{1,2m}=\frac{m}{2m+1},
\end{equation}%
which interestingly coincides with a subsequence of the Jain's series  given by 

\begin{equation}
\nu _{\mbox{\tiny Jain}}=\frac{m}{mk\pm 1}\qquad m,k/2=1,2,3,\ldots .
\label{jain}
\end{equation}
This  contains some  experimentally observed filling fractions. It has
been shown that the hierarchy scheme of states proposed by Jain \cite{Jain}
for the fractional quantum Hall effect can be viewed in terms of an
effective theory of composite fermions \cite{BW}. There the electrons are
thought of as dressed by magnetic fluxes leading to the filling factor. 
In type IIB superstring picture, it has been shown that fractional quantum
Hall states for the Jain filling factor can be related with the integer
quantum Hall states of the composite fermions. Using an effective field
theory description with Chern-Simons action, the Jain relation can be
interpreted as the result of the perturbative renormalization of the integer
QHE by the auxiliary heavy fermions. More details on this construction are
given in \cite{GKK}. However, here our analysis  is  based on hierarchy describtions 
 using Chern-Simons models dual to M-theory compactification on $ADE$ eight dimensional manifolds.  It should be interesting to find the connection between these two ways using string theory data. This will be addresed  elsewhere.

In the end of this subsection, we note  that  the integer $m$ appearing in the
Jain's series is related to the dimension of $H_{4}(M_{8},\mathbb{Z})$ for
the $A_{2m}$ geometry. At this level, one may ask the following question: Is
there any interpretation for the integer $k$?. In what follows, we speculate
on it. For this reason, it may be useful to introduce geometries with genus $%
g>0$. In fact, we will replace the zeroth Hirzebruch $F_{0}$ surface by $%
F_{0,g}=\Sigma _{g}\times \Sigma _{g}$ where $\Sigma _{g}$ is the Riemann
surface of genus $g$ that substitutes the $\mathbb{CP}^{1}$ sphere. Suppose,
for simplicity, that the cohomology space $H_{4}(M_{8},\mathbb{Z})$ is of
rank 2 and that $F_{0,g}^{1}$ and $F_{0,g}^{2}$ are complex surfaces
representing its generators $[F_{0,g}^{1}]$, $[F_{0,g}^{2}]$. The
intersection form is given by
\begin{equation}
\left(
\begin{array}{cc}
\lbrack F_{0,g}^{1}].[F_{0,g}^{1}] & [F_{0,g}^{1}].[F_{0,g}^{2}]\cr
[F^{2}_{0,g}].[F_{0,g}^{1}] & [F_{0,g}^{2}].[F_{0,g}^{2}]%
\end{array}%
\right) ,  \label{matrix}
\end{equation}%
where now $[F_{0,g}^{1}]\cdot \lbrack F_{0,g}^{2}]$ denotes the algebraic
intersection of $F_{0,g}^{1}$ and $F_{0,g}^{2}$. For any manifold $F_{0,g}$,
the geometry can be regarded as the zero section of the cotangent bundle
over $F_{0,g}$. Thus, the self intersection is equal to its minus Euler
number
\begin{equation*}
\lbrack F_{0,g}]\cdot \lbrack F_{0,g}]=2\Sigma _{g}\cdot \Sigma _{g}=2(2g-2).
\end{equation*}%
To get the intersection number between $F_{0,g}^{1}$ and $F_{0,g}^{2}$, one
may extend the result of the Riemann surfaces with genus 0 and 1. Assuming
that the complex surfaces meet negatively for $g>1$, we expect to have the
following intersection form
\begin{equation}
\lbrack F_{0,g}^{1}]\cdot \lbrack F_{0,g}^{2}]=2(1-g)
\end{equation}%
For more general extended geometries, one expects to have
\begin{equation}
\lbrack F_{0,g}^{i}]\cdot \lbrack F_{0,g}^{j}]=2(g-1)C^{ij}.
\end{equation}%
Now, specializing the computation to $A_{2(g-1)n}$, for $g>1 $,
quiver gauge theory for the vector charge $q_{i}=\delta _{i,p}$ with
$p=1$ or $n$, we get a filling factor coincides exactly with a
subsequence of Jain's series and the integer $k$ can be expressed in
terms of the genus as follows
\begin{equation}
k=2(g-1).
\end{equation}
which is, as it should be, even integer.
\subsection{$DE$ quiver gauge models}

We have mainly considered finite $A_n$ quivers. Here we would like to
generalize this to the case where the quiver gauge models are based on
finite $DE$ Dynkin diagrams. The general structure is quite like what we
have seen for $A_n$. First of all, in QHE, it is very crucial that the
filling fraction has odd-denominator (actually, there is no intrinsically
string theoretic understanding of the opposite). Admitting this fact, we see
that $D_{n}$ and  $E_{7}$ quiver gauge models do not provide \textquotedblleft
expected\textquotedblright\ results because the corresponding Cartan
matrices have even-determinant (4 and 2 respectively). As for the algebra $%
E_{8}$, for which the determinant is equal to 1, it always gives whole
filling factors. However, in the case of $\mbox{U(1)}^{6}$ quiver gauge
models based on finite $E_{6}$ Dynkin graph, we get interesting fractions.
For further illustration we display, in the table below, our calculations of
the filling factor $\nu $ concerning the $DE$ quiver gauge models for
typical values of the vector charge%
\begin{equation}
\begin{tabular}{lll}
\hline
& vector charge & $\nu $ \\ \hline\hline
$D_{n}$ & $(10\ldots 00)$ & 1/2 \\
& $(00\ldots 01)$ or $(0\ldots 010)$ & $n/8$ \\
$E_{6}$ & $(100000)$ or $(000010)$ & 2/3 \\
$E_{7}$ & $(0000010)$ & 3/4 \\
$E_{8}$ & $(00000010)$ & 1 \\ \hline
\end{tabular}%
\end{equation}%
The value $\nu =2/3$ associated with the $E_{6}$ case is more convincing
one. In fact, it belongs to the \textquotedblleft minus\textquotedblright\
subsequence of the Jain's series: $\nu =\frac{n}{2n-1}$ with $n=2$.

\section{Conclusion and comments}

In this letter, we have given M-theory derivation of FQHS using
3-dimensional Chern-Simons gauge theories based on Dynkin diagrams. This
construction, based on Lie\textrm{\ }algebras, leads to a general form of
the filling factor (\ref{factor}) and gives a unified description of several
kinds of FQH series including\ Laughlin, Haldane, Jain and hierarchical
ones.

Using M-theory/Type IIA duality, we have reproduced fractional values of the
filling factor observed experimentally using D6-branes wrapping a collection
of intersecting $F_{0}$ geometries according to Dynkin diagrams of finite $%
A_{n}$ type algebras. In particular, the Jain's series can be recovered by
considering quiver gauge theory based on $A_{2m}$ Dynkin diagrams. However,
for $n=2m+1$ we get $\nu =1-\frac{1}{m+1}$ which coincides with known
filling fractions in the literature \cite{wen}. We have also analyzed the
finite $DE$ quiver gauge theories.

Our approach is adaptable to a broad variety of geometries whose
intersection forms may be represented by extended Cartan matrices.
We intend to discuss elsewhere the extension of this explicit study
to the indefinite extended geometries as well as those brane
realizations based on D6-branes wrapping del Pezzos $dP_{n}$\ with
$1<n\leq 8$\ where $F_{0}$\ is just the leading
$dP_{1}$\textrm{\textrm{.} }In connection with that, it would
therefore be of interest to consider the two following:
\newline (\textbf{1}) use the tools developed recently in the framewok of
F-theory-GUT to construct new classes of QHS that are embedded in
M-theory on CY4-folds. Obviously, the physics between the two topics
is different; the role played by the 7-brane wrapping del Pezzos in
F-theory GUT is now played by 6-brane wrapping the same 4-cycles. In
other words, in both F-theory-GUT and QHS in M-theory, we have the
same kind of CY4-folds. \newline (\textbf{2}) choose a given FQHE
series $\nu $, and solve eq(\ref{factor}) to end with a matrix which is not necessary of
Cartan type. We then expect the existence of models that are
associated with Borcherds symmetries giving one more evidence for
these kinds of M-theory symmetries. For a connection between
Borcherds and M-theory, see \cite{DAM} and refs therein. We believe
that these issues deserve to be studied further.

\emph{Acknowledgments.} This work is supported by the program Protars III
D12/25. AS is supported by CICYT (grant FPA-2006-02315 and grant
FPA-2009-09638) and DGIID-DGA (grant 2007-E24/2). We thank also the support
by grant A/93357/07 and A/024147/09.

\end{document}